\documentstyle[pra,aps,twocolumn,psfig,floats,amssymb]{revtex}

\def\j{}

\begin{document}

\wideabs{
\title{Pulse-order invariance of the initial-state population
in multistate chains driven by delayed laser pulses}
\author{N. V. Vitanov}
\address{Helsinki Institute of Physics, P. O. Box 9,
 00014 Helsingin yliopisto, Finland}
\date{\today }
\maketitle
\begin{abstract}
This paper shows that under certain symmetry conditions the probability
of remaining in the initial state (the probability of no transition)
in a chainwise-connected multistate system driven by two or more delayed
laser pulses does not depend on the pulse order.
\end{abstract}
}

The process of stimulated Raman adiabatic passage (STIRAP)
has received a great deal of attention in the past decade
\cite{Bergmann95,Bergmann98} because of its potential
for efficient and robust population transfer between two states
$\psi_1$ and $\psi_3$ via an intermediate state $\psi_2$.
STIRAP uses two delayed laser pulses, a pump pulse $\Omega_P(t)$
linking states $\psi_1$ and $\psi_2$ and a Stokes
pulse $\Omega_S(t)$ linking states $\psi_2$ and $\psi_3$.
By applying the Stokes pulse before the pump pulse (counterintuitive
order) and maintaining adiabatic-evolution conditions and two-photon
resonance between states $\psi_1$ and $\psi_3$, one ensures
complete and smooth transfer of population from $\psi_1$ to
$\psi_3$, regardless of whether the intermediate state
is on or off single-photon resonance.
Applying the two pulses in the intuitive order
[$\Omega_P(t)$ before $\Omega_S(t)$] leads to oscillations
in the on-resonance case
and to STIRAP-like transfer in the off-resonance case.
The success of STIRAP has prompted its extension to
multistate chainwise-connected systems
\cite{Shore91,Pillet93,Goldner94,Theuer98,Vitanov98multi,Vitanov98tune},
where a similar distinction between the intuitive
and counterintuitive pulse orders exists.

In view of the great difference in the {\em final-state} population for
the two pulse orders, surprisingly, the {\em initial-state} population
has been found to be the same for both orders in the three-state case,
provided the Hamiltonian has a certain symmetry \cite{Vitanov97}.
The present paper extends this result to {\em multistate} chains.
Thus it establishes another similarity between three-state
and multistate systems.

The time evolutions of the probability amplitudes
${\bf c}(t)=\left[c_{1}(t),c_{2}(t),\ldots ,c_{N}(t)\right] ^{T}$
of the $N$ states satisfy the Schr\"{o}dinger equation
(in units $\hbar =1$) \cite{Shore90}, 
\begin{equation}
i {\bf \dot c}(t)={\bf H}(t){\bf c}(t).  \label{3SEqs}
\end{equation}
In the rotating-wave approximation the Hamiltonian of the
multistate chain is given by the tridiagonal matrix 
\begin{equation}
{\bf H} = \left[ 
\begin{array}{cccccc}
0 & \Omega _{12} & 0 & \vdots  & 0 & 0 \\ 
\Omega _{12} & \Delta _{2} & \Omega _{23} & \vdots  & 0 & 0 \\ 
0 & \Omega _{23} & \Delta _{3} & \vdots  & 0 & 0 \\ 
\cdots  & \cdots  & \cdots  & \ddots  & \cdots  & \cdots  \\ 
0 & 0 & 0 & \vdots  & \Delta _{N-1} & \Omega _{N-1,N} \\ 
0 & 0 & 0 & \vdots  & \Omega _{N-1,N} & 0
\end{array}
\right] .  \label{H}
\end{equation}
The system is supposed to have $N=2n+1$ states
and the Rabi frequencies $\Omega_{j,j+1}(t)$ obey the relations 
\begin{mathletters}\label{pulses}
\begin{eqnarray}
\Omega_{j,j+1}(t) &=& \left\{ \begin{array}{ll}
\xi_j \Omega_P(t), & j\text{ odd}, \\ 
\xi_j \Omega_S(t), & j\text{ even},
\end{array}\right.   \\
\xi_j &=& \xi_{N+1-j}, \\
\Omega_P(t) &=& \Omega_0 f(t-\tau),\\ 
\Omega_S(t) &=& \Omega_0 f(t+\tau),
\end{eqnarray}
\end{mathletters}
and $f(-x)=f(x)$.
The functions $\Omega_P(t)$ and $\Omega_S(t)$ describe the
envelopes of the two pulses, $2\tau$ is the pulse delay,
$\Omega_0$ is an appropriate unit of Rabi frequency,
and the (constant) relative coupling strengths $\xi_j$
are proportional to the corresponding Clebsch-Gordan coefficients.
The detunings are supposed to obey the relations
\begin{mathletters} \label{detunings}
\begin{eqnarray}
\Delta_j(t) &=& \Delta_{N+1-j}(t), \\
\Delta_j(t) &=& \Delta_j(-t),\qquad (j = 2,3,\ldots ,n+1).
\end{eqnarray}
\end{mathletters}
For example, Eqs. (\ref{pulses}) and (\ref{detunings}) apply to
$2J+1$-state systems ($J$ integer), formed by the sublevels in
$J \rightarrow J$ or $J \rightarrow J-1$ transition, coupled by
two laser pulses $\Omega_P(t)$ and $\Omega_S(t)$
with $\sigma^+$ and $\sigma^-$ polarizations
\cite{Pillet93,Goldner94,Theuer98}.

I shall show that when conditions (\ref{pulses}) and (\ref{detunings})
are satisfied the probability of remaining in the initial state
(the probability of no transition) does not depend on the pulse order,
i.e., it is invariant upon the interchange of
$\Omega_P(t)$ and $\Omega_S(t)$.
Since the $\Omega_P\rightleftarrows \Omega_S$ swap is equivalent
to the index change $j\rightarrow N+1-j$ in ${\bf H}(t)$,
the $\Omega_P\rightleftarrows \Omega_S$ invariance
of the population of the initial state $\psi_j$
is equivalent to the assertion that for a given pulse order,
the probability of remaining in state $\psi_j$,
provided the system is initially in state $\psi_j$,
is equal to the probability of remaining in state $\psi_{N+1-j}$,
provided the system is initially in state $\psi_{N+1-j}$.
In terms of the transition matrix ${\bf U}(+\infty,-\infty)$,
defined by
${\bf c}(+\infty)={\bf U}(+\infty,-\infty) {\bf c}(-\infty)$,
this invariance means that for any $j = 1,2,\ldots,n+1$,
\begin{equation}
U_{jj}(+\infty,-\infty) = U_{N+1-j,N+1-j}(+\infty,-\infty).
\label{Ujj}
\end{equation}

The proof of Eq.~(\ref{Ujj}) is carried out in several steps.
The first step is to show that the eigenvalues and the eigenstates
of ${\bf H}(t)$ have certain symmetric properties.
These properties lead to symmetries of the Hamiltonian
in the adiabatic basis, which determine certain symmetries
of the adiabatic transition matrix, which in turn lead to
the property (\ref{Ujj}) of the diabatic transition matrix.

It follows from Eqs. (\ref{pulses}) and (\ref{detunings}) that the
$\Omega_P\rightleftarrows \Omega_S$ swap is equivalent to
time reversal in ${\bf H}(t)$, 
\begin{equation}
\Omega_{P}(t)\rightleftarrows \Omega_S(t)\qquad
\text{is equivalent to\qquad } t\rightarrow -t.
\label{TimeReversal}
\end{equation}
Hence, since the $\Omega_P\rightleftarrows \Omega_S$ swap
does not change the eigenvalues of the Hamiltonian,
${\bf H}(-t)$ has the same eigenvalues as ${\bf H}(t)$.
The eigenvalues $\lambda_j(t)$ of ${\bf H}(t)$ are therefore even
functions of time, 
\begin{equation}
\lambda_j(-t)=\lambda_j(t),\qquad (j=1,2,\ldots ,N).
\label{lambda}
\end{equation}
Since ${\bf H}(t)$ is real and symmetric, its eigenvalues are real
and its eigenstates can be chosen real too.
The components of the eigenstates (the adiabatic states)
${\bf w}^j(t) = [w_1^j(t),w_2^j(t),\ldots ,w_N^j(t)]^T$
are expressed in terms of $w_1(t)$
(for simplicity, the label $j$ is omitted for the moment) as
\begin{eqnarray*}
\frac{w_2\j(t)}{w_1\j(t)} &=& \frac{\lambda(t)}{\xi_1\Omega_P(t)}
	\equiv g_2(t), \\
\frac{w_3\j(t)}{w_1\j(t)} &=& \frac{\lambda(t)
 \left[ \lambda(t)-\Delta_2(t)\right]
 -\xi_1^2\Omega_P^2(t)}{\xi_1\xi_2\Omega_P(t)\Omega_S(t)}
	\equiv g_3(t), \\
&&\ldots ,
\end{eqnarray*}
and in terms of $w_N\j(t)$ as 
\begin{eqnarray*}
\frac{w_{N-1}\j(t)}{w_N\j(t)} &=&\frac{\lambda(t)}{\xi_1\Omega_S(t)}
	\equiv g_2(-t), \\
\frac{w_{N-2}\j(t)}{w_N\j(t)} &=&\frac{\lambda(t)
 \left[ \lambda(t)-\Delta_2(t)\right]
 -\xi_1^2\Omega_S^2(t)}{\xi_1\xi_2\Omega_P(t)\Omega_S(t)}
	\equiv g_3(-t), \\
&&\ldots 
\end{eqnarray*}
Generally, one can write $w_k\j(t)/w_1\j(t)=g_k(t)$ and
$w_{N+1-k}\j(t)/w_N\j(t)=g_k(-t)$.
For $k=n+1$, one finds $g_{n+1}(-t)w_N\j(t)=g_{n+1}(t)w_1\j(t)$.
It follows that
\[
\frac{w_{N+1-k}\j(t)}{w_k\j(t)}
= \frac{g_k(-t)}{g_k(t)}\frac{g_{n+1}(t)}{g_{n+1}(-t)},
\]
for any $k=1,2,\ldots ,n+1$. Hence
\begin{eqnarray}
w_1\j(t) &=& g_{n+1}(-t) / \nu (t),  \nonumber \\
w_2\j(t) &=& g_{2}(t)g_{n+1}(-t) / \nu (t),  \nonumber \\
&&\cdots,   \nonumber \\
w_{n+1}\j(t) &=& g_{n+1}(t)g_{n+1}(-t) / \nu (t),  \nonumber \\
&&\cdots,   \nonumber \\
w_{N-1}\j(t) &=& g_{n+1}(t)g_{2}(-t) / \nu (t),  \nonumber \\
w_N\j(t) &=& g_{n+1}(t) / \nu (t).
\label{w_k}
\end{eqnarray}
The normalization factor $\nu (t)$ is obviously invariant upon
time reversal, which means that $\nu (-t)=\nu (t)$.
Equations (\ref{w_k}), which are valid for $g_{n+1}^j(t)\neq 0$
(case I), lead to the relation (with the label $j$ restored)
\begin{mathletters}
\label{v(-t)}
\begin{equation}
w_k^j(-t)=w_{N+1-k}^j(t),\qquad \text{(case I)},
\label{v(-t)even}
\end{equation}
with $k=1,2,\ldots ,n+1$.

If $g_{n+1}^m(t)=0$ (case II) for a certain $\lambda_m(t)$, we have
$w_{n+1}^m(t)=0$ and $w_{n+2}^m(-t)=-w_{n}^m(t)$, which leads to 
\begin{equation}
w_k^m(-t)=-w_{N+1-k}^m(t), \qquad \text{(case II)},
\label{v(-t)odd}
\end{equation}
\end{mathletters}
with $k=1,2,\ldots ,n+1$.
Such a case arises for the zero-eigenvalue eigenstate in systems with
$N=3,7,11,\ldots $ states and zero detunings.

The symmetry relations (\ref{v(-t)}) for the adiabatic states determine
certain symmetries of the Hamiltonian in the adiabatic basis.
The transformation from the original (diabatic) basis to the adiabatic basis,
${\bf c}(t)={\bf W}(t){\bf a}(t)$,
is carried out by the orthogonal matrix ${\bf W}(t)$,
whose columns are the normalized eigenvectors ${\bf w}^j(t)$.
Here ${\bf a}(t)=\left[ a_1(t),a_2(t),\ldots,a_N(t)\right]^T$
is the column-vector of the adiabatic probability amplitudes.
The Schr\"{o}dinger equation in the adiabatic basis reads 
\begin{equation}
i{\bf \dot a}(t)={\bf H}^a(t){\bf a}(t),
\label{SEq-adb}
\end{equation}
where ${\bf H}^a(t)={\bf H}^{\text{adb}}(t)+{\bf H}^{\text{nonadb}}(t)$
with
\begin{mathletters}
\begin{eqnarray}
{\bf H}^{\text{adb}}(t) &=& {\bf W}^T(t){\bf H}(t){\bf W}(t), \\
{\bf H}^{\text{nonadb}}(t) &=& -i{\bf W}^T(t){\bf \dot W}(t).
\end{eqnarray}
\end{mathletters}
The adiabatic part ${\bf H}^{\text{adb}}(t)$ is a diagonal matrix
containing the eigenvalues $\lambda_j(t)$ of ${\bf H}(t)$
on the main diagonal.
The nonadiabatic part ${\bf H}^{\text{nonadb}}(t)$ has zeros
on the main diagonal, while the off-diagonal elements are
equal to the nonadiabatic couplings
$H_{jk}^{\text{nonadb}}(t)=-i{\bf w}^j(t)\cdot {\bf \dot w}^k(t)$.
It is readily seen from Eq.~(\ref{v(-t)even}) that the nonadiabatic
coupling between two case-I adiabatic states ${\bf w}^j (t)$ and
${\bf w}^k (t)$ is an odd function of time.
Really, 
\begin{mathletters}
\begin{eqnarray}
H_{jk}^{\text{nonadb}}(-t) &=&i\sum_{l=1}^N w_l^j(-t)\dot w_l^k(-t)
\nonumber \\
&=&i\sum_{l=1}^N w_{N+1-l}^j(t) \dot w_{N+1-l}^k(t) \nonumber\\
&=&-H_{jk}^{\text{nonadb}}(t),\ \ \text{(case I $\cdot$ case I)}.
\label{caseA}
\end{eqnarray}
The nonadiabatic coupling between a case-I eigenstate ${\bf w}^j(t)$
and a case-II eigenstate ${\bf w}^m(t)$ is an even function,
\begin{equation}
H_{jm}^{\text{nonadb}}(-t) = H_{jm}^{\text{nonadb}}(t),
	\ \ \text{(case I $\cdot$ case II)}.
\label{caseB}
\end{equation}
\end{mathletters}

The symmetry of ${\bf H}^a(t)$ determines a certain symmetry of the
adiabatic transition matrix ${\bf U}^a(+\infty,-\infty)$, defined
as  ${\bf a}(+\infty)={\bf U}^a(+\infty,-\infty){\bf a}(-\infty)$.
In order to find it, I introduce the evolution matrix ${\bf G}(t,0)$
via ${\bf a}(t)={\bf G}(t,0){\bf a}(0)$.
Evidently, the first column of ${\bf G}(t,0)$ is the solution of 
Eq.~(\ref{SEq-adb}) for the initial condition
${\bf a}(0) = \left( 1,0,0,\ldots,0\right)^T$,
the second column is the solution for
${\bf a}(0) = \left( 0,1,0,\ldots ,0\right)^T$, and so on.
When all nonadiabatic couplings are odd functions of time
[Eq.~(\ref{caseA})], time reversal in Eq.~(\ref{SEq-adb})
is equivalent to complex conjugation of ${\bf a}(t)$ (case A).
When a case-II eigenstate ${\bf w}^m(t)$ exists [then the nonadiabatic
couplings involving it are even functions, Eq.~(\ref{caseB})],
time reversal in Eq.~(\ref{SEq-adb}) is equivalent to complex
conjugation of ${\bf a}(t)$ and change of sign of $a_m(t)$ (case B).
This means that
\begin{equation}\label{G}
{\bf G}(-t,0)=\left\{ 
\begin{array}{ll}
{\bf G}^{*}(t,0), & \text{(case A)}, \\ 
{\bf IG}^{*}(t,0){\bf I}, & \text{(case B)},
\end{array}
\right. 
\end{equation}
where ${\bf I}$ is a diagonal matrix with units on its diagonal,
except the $(m,m)$-th element which is $-1$.
It follows from Eq.~(\ref{G}) and the unitarity of ${\bf G}$ that 
\begin{eqnarray*}
{\bf U}^a(+\infty ,-\infty )
&=&{\bf G}(+\infty ,0){\bf G}(0,-\infty )\\
&=&{\bf G}(+\infty ,0){\bf G}^{\dagger }(-\infty ,0) \\
&=&\left\{ 
\begin{array}{ll}
{\bf G}(+\infty ,0){\bf G}^{T}(+\infty ,0), & \text{(case A)}, \\ 
{\bf G}(+\infty ,0){\bf IG}^{T}(+\infty ,0){\bf I}, & \text{(case B)}.
\end{array}\right. 
\end{eqnarray*}
Hence 
\begin{equation}\label{U}
\left[ {\bf U}^a(+\infty,-\infty)\right]^T = \left\{ 
\begin{array}{ll}
{\bf U}^a(+\infty,-\infty), & \text{(case A)}, \\ 
{\bf IU}^a(+\infty,-\infty){\bf I}, & \text{(case B)}.
\end{array}
\right. 
\end{equation}

The transition matrices in the diabatic and adiabatic bases
are related by 
\[
{\bf U}(+\infty,-\infty)
 = {\bf W}(+\infty)
   {\bf U}^{a}(+\infty,-\infty)
   {\bf W}^T(-\infty). 
\]
Then one finds from Eqs.~(\ref{v(-t)}) and (\ref{U})
that in both cases A and B,
\begin{eqnarray*}
&&U_{N+1-j,N+1-j}(+\infty,-\infty)\\
&&\qquad = \sum_{k,l=1}^N
 w_{N+1-j}^{k}(+\infty)
 U_{kl}^{a}(+\infty,-\infty)
 w_{N+1-j}^{l}(-\infty) \\
&&\qquad = \sum_{k,l=1}^N
 w_j^k(-\infty) U_{lk}^a(+\infty,-\infty) w_{j}^{l}(+\infty) \\
&&\qquad = U_{jj}(+\infty,-\infty).
\end{eqnarray*}
This completes the proof.


\begin{figure}[tb]
\centerline{\psfig{width=80mm,file=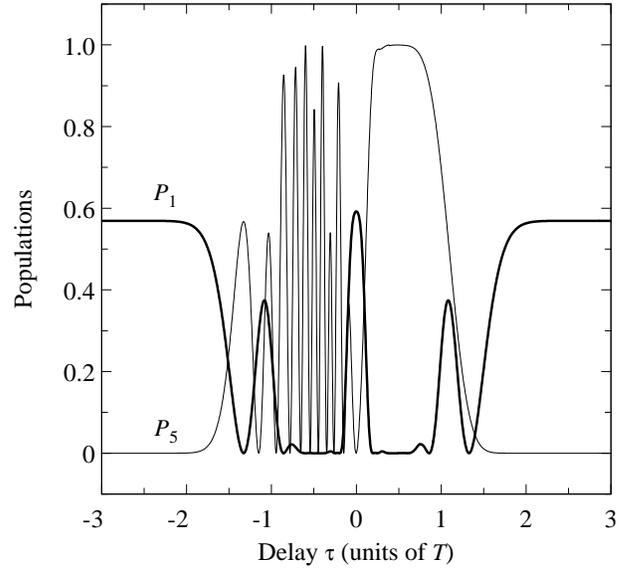}}
\vspace*{2mm}
\caption{
The initial-state population $P_1$ and the final-state population $P_5$
for a five-state system, initially in state $\psi_1$, plotted
against the pulse delay $\tau$ in the resonance case
($\Delta_2 = \Delta_3 = \Delta_4 = 0$).
The Rabi frequencies of the two pulses are given
by Eqs.~(\protect\ref{pulses}) and have Gaussian shapes,
$\Omega_P(t)=\Omega_0\exp[-(t-\tau)^2/T^2]$ and
$\Omega_S(t)=\Omega_0\exp[-(t+\tau)^2/T^2]$,
with $\xi_1=\xi_4=\protect\sqrt{1/3}$, $\xi_2=\xi_3=\protect\sqrt{1/2}$,
and $\Omega_0 T =30$.
}
\label{Fig}
\end{figure}


It should be emphasized that the pulse-order invariance applies to the
population of the initial state only, while the populations
of all other states depend on the pulse order.
This is clearly demonstrated in Fig. \ref{Fig} where
the initial-state population $P_1$ and the final-state population $P_5$
are plotted against the pulse delay $\tau$ in the case of
a five-state system, initially in state $\psi_1$.
The figure shows that $P_5$ behaves very similarly to STIRAP
with a broad plateau of high transfer efficiency for $\tau > 0$
and oscillations for $\tau < 0$
\cite{Bergmann95,Bergmann98,Vitanov97}.
In contrast, $P_1$ is a symmetric function of $\tau$,
as follows from the above results.

Finally, the pulse-order invariance of the initial-state population
has been derived without the assumption of adiabatic evolution.
Hence it applies to the general nonadiabatic case, as long as the pulse
duration is long enough to validate the rotating-wave approximation.

This work has been supported financially by the Academy of Finland.

\end{document}